# Characterization of InSb quantum wells with atomic layer deposited gate dielectrics

M. M. Uddin,[1] H. W. Liu,[2,3,a)] K. F. Yang,[2] K. Nagase,[2] T. D. Mishima,[4] M. B. Santos,[4] and Y. Hirayama[1,2,b)]

[1]*Department of Physics, Tohoku University, Sendai, Miyagi 980-8578, Japan*

[2]*ERATO Nuclear Spin Electronics Project, Sendai, Miyagi 980-8578, Japan*

[3]*State Key Laboratory of Superhard Materials and Institute of Atomic and Molecular Physics, Jilin University, Changchun 130012, People's Republic of China*

[4]*Homer L. Dodge Department of Physics and Astronomy, University of Oklahoma, 440 West Brooks, Norman, Oklahoma 73019-2061, USA*

We report magnetotransport measurements of a gated InSb quantum well (QW) with high quality $Al_2O_3$ dielectrics (40 nm thick) grown by atomic layer deposition. The magnetoresistance data demonstrate a parallel conduction channel in the sample at zero gate voltage ($V_g$). A good interface between $Al_2O_3$ and the top InSb layer ensures that the parallel channel is depleted at negative $V_g$ and the density of two-dimensional electrons in the QW is tuned by $V_g$ with a large ratio of $6.5 \times 10^{14}$ $m^{-2}V^{-1}$ but saturates at large negative $V_g$. These findings are closely related to layer structures of the QW as suggested by self-consistent Schrödinger-Poisson simulation and two-carrier model.

[a)]Electronic email: liuhw@ncspin.jst.go.jp

[b)]Electronic email: hirayama@m.tohoku.ac.jp

The typical narrow-gap semiconductor InSb has attracted much attention for next-generation high-speed electronics owing to its unique material properties of large dielectric constant (16.8), small effective mass (0.014 $m_e$, $m_e$ is the free-electron mass), and high room-temperature mobility (77000 cm$^2$/Vs).[1-3] More recently, a two-dimensional electron gas (2DEG) confined in single InSb quantum wells (QWs) has emerged as a promising candidate in the research field of spintronics by virtue of either strong spin-orbit coupling or a giant Landé $g$-factor.[4-6] Further studies including implementation of electron-spin precession in a spin field-effect transistor[7] and nuclear-spin coherence in a nanoscale region (e.g., quantum point contact)[8] require a gated InSb QW. Up to now, however, fabrication of the gated InSb QW remains challenging: either comparatively large conductance and high-density dislocation of an $Al_xIn_{1-x}Sb$ buffer layer between the QW and the substrate or relatively low Schottky barrier heights[9] hinder the application of back gating or Schottky contact to the InSb QW. Because of these difficulties, we here focus on fabricating the gated InSb QW using top gating approach with gate dielectrics.

Atomic layer deposition (ALD) has been extensively used to grow high quality high-k dielectrics on quantum structures of III-V compound semiconductors (GaAs, InAs, GaN, GaSb, etc.).[10] However, so far the study of the InSb QW with ALD gate dielectric is quite limited probably because of the problem in interfacing high-k/(Al)InSb.[9] In this work, we present a gated InSb QW with high quality ALD-$Al_2O_3$ dielectric (40 nm thick). Unlike in the work of Brask *et al.* [9], where bridging atoms are embodied at the high-k/$Al_xIn_{1-x}Sb$ interface, we *directly* deposit $Al_2O_3$ on the top InSb layer of the QW. A rapid change in the 2DEG density in response to gate voltages has been demonstrated, suggesting that the Fermi level at the $Al_2O_3$/InSb interface is weakly pinned. This allows us to address another issue that attracts less attention: the magnetoresistive (MR) property of the gated InSb QW strongly depends on layer structures of the QW, which has been elucidated by self-consistent Schrödinger-Poisson (SP) simulation and the two-carrier model.

Figure 1(a) shows a cross section of the InSb QW with ALD-$Al_2O_3$ dielectric. The InSb/$Al_xIn_{1-x}Sb$ multilayers were grown on GaAs (001) substrates using an Intevac Modular Gen II molecular beam epitaxy system.[11] Two silicon (Si) δ-doped layers with the same donor density of $3.6 \times 10^{15}$ m$^{-2}$ (dashed line) are located above the QW: one is placed 15 nm above the QW to supply electrons in the well and the other 30 nm away from the top InSb layer to provide electrons to the surface state. A Hall bar pattern (80 × 30 μm$^2$) was delineated by mesa etching and indium was employed as ohmic contacts (not shown) using electron-beam deposition and lift-off. After being washed with acetone, Isopropyl and purified water and dried with nitrogen ($N_2$) gas, the InSb Hall bar was coated with a lift-off pattern of positive photoresist (S1813) and loaded into the ALD system (Savannah S-100, Cambridge Nano Tech. Inc.). The $Al_2O_3$ dielectric was grown at 150°C using trimethylaluminum (TMA) with water ($H_2O$) and ozone ($O_3$) as the oxygen source. This growth temperature also ensures successful lift-off of the photoresist S1813. The first ten reaction cycles using alternate micropulses of $H_2O$ (0.015s)/$N_2$ (waiting time: 20s) and TMA (0.015s)/$N_2$ (waiting time: 20s) were set to deposit ~ 1nm of $Al_2O_3$ on the sample to prevent the photoresist pattern being damaged by $O_3$ in the latter process. An *in situ* surface treatment was then performed in 20 subsequent cycles using $O_3$ (0.025s)/$N_2$ pulses (waiting time: 15s). Finally, 363 cycles using alternate micropulses of TMA (0.015s)/$N_2$ (waiting time: 20s) and $O_3$ (0.05s)/$N_2$ (waiting time: 15s) were taken under the pressure of 1 Torr. The deposited $Al_2O_3$ film is about 40 nm thick with a small root-mean-square roughness[12] of ~ 0.5 nm as seen by atomic force microscopy (AFM) (inset of Fig. 1(b)), indicating a smooth and dense surface. Indium is used as top-gate metal. Figure 1(b) shows the current-gate voltage ($V_g$) characteristic of the $Al_2O_3$ film at 2 K. The current is measured between top-gate and Ohmic contact to the QW. Current leakage is clearly prominent at large negative $V_g$, suggesting high quality of this dielectric. Note that, at positive bias, an electron channel is induced around the surface of a Hall bar covered by $Al_2O_3$ and is connected to the 2DEG via side wall contact. In this case, current leakage might occur even at a small bias provided the breakdown property of the thinner $Al_2O_3$ film across the Hall bar edge (circle mark in Fig. 1(a)) deteriorates. This problem can be avoided by a gate air-bridge technique.[13]

The longitudinal resistivity $\rho_{xx}$ and Hall resistivity $\rho_{xy}$ of the gated InSb QW vs. magnetic field ($B$) at different $V_g$ are shown in Fig. 2(a). The MR plot at $V_g = 0$ (thick line) exhibits an increasing resistivity background in $\rho_{xx}$ and a sublinear field response in $\rho_{xy}$, which is very different from the standard MR property of single 2DEGs featured in Hall plateau with zero $\rho_{xx}$ at high $B$ [e.g., the cases of $V_g$ = -1.6 V (thin line) and the InSb QW before the ALD process (inset)]. This observation reminds us that a parallel conduction channel is formed in the sample after the ALD process. According to a two-carrier model,[14] the sublinear $\rho_{xy}$-$B$ data can be fit well by $\rho_{xy} = \frac{\sigma_{xy}}{\sigma_{xx}^2 + \sigma_{xy}^2}$ (dashed line), where $\sigma_{xx} = \frac{en_p\mu_p}{1+(\mu_p B)^2} + \frac{en_{QW}\mu_{QW}}{1+(\mu_{QW}B)^2}$, $\sigma_{xy} = \frac{en_p B\mu_p^2}{1+(\mu_p B)^2} + \frac{en_{QW} B\mu_{QW}^2}{1+(\mu_{QW}B)^2}$, and $n_{p(QW)}$ and $\mu_{p(QW)}$ are the density and mobility of electrons in the parallel channel (in the QW), respectively. The fit yields $n_p = 3.64 \times 10^{15}$ m$^{-2}$, $\mu_p = 0.53$ m$^2$/Vs, and $\mu_{QW} = 17.5$ m$^2$/Vs using the parameter value of $n_{QW} = 3.86 \times 10^{15}$ m$^{-2}$ obtained from the fast Fourier transform (FFT) analysis of Shubnikov-de-Haas (SdH) oscillations in $\rho_{xx}$ at $V_g = 0$. The parallel channel is found to be fully depleted by applying large negative $V_g$ as reflected by the MR property of the single 2DEG (e.g., the case of $V_g = -1.6$ V in Fig. 2(a)). The gate-dependent $n_{QW}$ and $\mu_{QW}$ are shown in Fig. 2(b), from which a relatively large ratio of d$n_{QW}$/d$V_g = 6.5 \times 10^{14}$ m$^{-2}$V$^{-1}$ (estimated in the region of -1.6 V $\leq V_g \leq$ 0.5 V) is obtained. This plot demonstrates that gate control of the InSb 2DEG works well.

The above MR property can be qualitatively understood by solving the SP equation self-consistently. Note that in this simulation a metal-insulator-semiconductor contact is treated as a Schottky contact.[15] The Schottky barrier height $\phi_B$ determined by the property of the InSb surface and/or the Al$_2$O$_3$/InSb interface corresponds to the energy difference between the conduction band minimum (CBM) $E_c$ of InSb and the Fermi level $E_F$ under thermal equilibrium conditions. The CB edge profile of the InSb QW and electron density distribution ($|\Psi|^2$) in the QW calculated using the reported material parameters[16] are illustrated in Fig. 3, where $E_F$ is at 0 eV and $E_c$ at zero depth is equal to $\phi_B$. The CB edge profile at $\phi_B \sim 0.128$ eV (thin solid line) is assigned to explain the MR property of the InSb QW before the ALD process (inset of Fig. 2(a)). Because only the CB edge of the QW is below $E_F$, a single conduction channel (i.e., the 2DEG) exists in the sample. Note that $\phi_B$ is set to be larger than the mid-gap value (0.118 eV at 2 K) of InSb based on the fact that the Fermi level is located above charge neutrality levels at the InSb surface[17], and hence the CB edge bends upwards after thermal equilibrium. As $\phi_B$ decreases (thick line in Fig. 3), the CB discontinuity at the interface between the two Al$_x$In$_{1-x}$Sb layers (circle) lies below $E_F$ and an additional electron channel emerges. This accounts for the parallel conduction channel formed in the InSb QW after the ALD process. The reduction of $\phi_B$ probably originates from the compensation of intrinsic acceptor surface states[17] and/or from the electric dipole effect[18,19] at the Al$_2$O$_3$/InSb interface. The reduced $\phi_B$ also lowers the CB edge of the QW, resulting in an increase of $n_{QW}$. This is consistent with the experimental result showing that $n_{QW}$ increases from $2.8 \times 10^{15}$ m$^{-2}$ to $3.86 \times 10^{15}$ m$^{-2}$ after the ALD process. A negative-bias potential effectively increases $\phi_B$ and raises the CB discontinuity entirely above $E_F$ (dashed line in Fig. 3). As a result, the parallel channel is depleted and the standard MR property of single 2DEGs reappear (the case of $V_g = -1.6$ V in Fig. 2(a)). Since the CB edge of the QW moves to higher energy at negative bias, $n_{QW}$ decreases as is shown in Fig. 2(b). At more negative bias, $n_{QW}$ becomes saturated because the valence band maximum (VBM) of InSb reaches the Fermi level and holes accumulate at the surface to screen the gate potential to some degree.[20] That is, the 2DEG in this sample cannot be completely depleted even by applying large negative $V_g$.

It is understood from the above discussion that $\phi_B$ is modulated by $V_g$ in a wide energy range, indicating that the Fermi level at the $Al_2O_3$/InSb interface is weakly pinned. Since the band gap of InSb and $Al_{0.1}In_{0.9}Sb$ in our sample is quite small, the Fermi level easily reaches the CBM or VBM of these materials and thereby affects the MR property as described above. Because of this, it is expected that the InSb QW with a single layer of $Al_xIn_{1-x}Sb$ above the QW is more appropriate for gate control of the InSb 2DEG provided that the component x of $Al_xIn_{1-x}Sb$ is chosen properly by taking account of the band gap value (linear in $x$[21]) and the lattice mismatch to InSb. In fact, Gilbertson *et al.* have employed the InSb QW with a single $Al_{0.15}In_{0.85}Sb$ layer above the QW for the study of gate control of the 2DEG but using sputtered $SiO_2$ as gate dielectric.[4] Such a QW structure together with the advantage of ALD in tailing the dielectric/semiconductor interface is believed to ensure complete depletion of the InSb 2DEG under gate control.

In conclusion, the MR property of a gated InSb QW with high quality ALD-$Al_2O_3$ dielectric has been characterized at 2 K and discussed using the SP simulation and the two-carrier model. Our results clearly demonstrate that both interface and QW layer structure issues are crucial for gate control of the InSb 2DEG. This work paves the way for further developments in top-down fabrication of InSb nanodevices.

The authors acknowledge K. Hashimoto for helpful discussions on AFM measurements. M.M.U. would like to thank IIARE, Tohoku University for S-SDC fellowship. H.W.L. thanks the Program for New Century Excellent Talents of the University in China.

**FIG. 1.** (Color online) (a) Cross section of a gated InSb QW with ALD-Al$_2$O$_3$ dielectric. The Si δ-doped layer is indicated by a dashed line and the area of ALD-Al$_2$O$_3$ at the corner of Hall bars is marked by a circle. (b) Gate leakage current vs. gate voltage $V_g$ at 2 K. Inset shows an AFM image of a 40 nm thick ALD-Al$_2$O$_3$ film grown at 150°C. The scale bar represents 500 nm.

**FIG. 2.** (Color online) (a) Magnetic-field ($B$) dependent longitudinal resistivity $\rho_{xx}$ and Hall resistivity $\rho_{xy}$ of the InSb QW at $V_g = 0$ V (thick line) and $V_g = -1.6$V (thin line). The dashed line is a fit for the nonlinear $\rho_{xy}$-$B$ plot calculated by a two-carrier model. Inset: $\rho_{xx}$ and $\rho_{xy}$ vs. $B$ for the InSb QW before the ALD process. As a comparison, the dotted line guides a linear dependence of $\rho_{xy}$-$B$. All the above measurements were performed at 2 K using a standard lock-in technique with a frequency of 13.3 Hz and an AC current of 35 nA. (b) Electron density ($n_{QW}$) and mobility ($\mu_{QW}$) of the 2DEG confined in the InSb QW as a function of $V_g$. Note that the data at $V_g > -1.2$ V where a parallel conduction channel is formed in the sample were calculated from the FFT analysis of SdH oscillations in $\rho_{xx}$ and the others from the classical Hall measurement.

**FIG. 3.** (Color online) Conduction-band profiles and the square of electron wave functions of the InSb QW obtained from self-consistent Schrödinger-Poisson solutions. The corresponding layer structures are indicated along the bottom axis. The band gap energy of all the layers is calculated at 2 K. The circle accentuates the CB discontinuity below the Fermi level $E_F$.

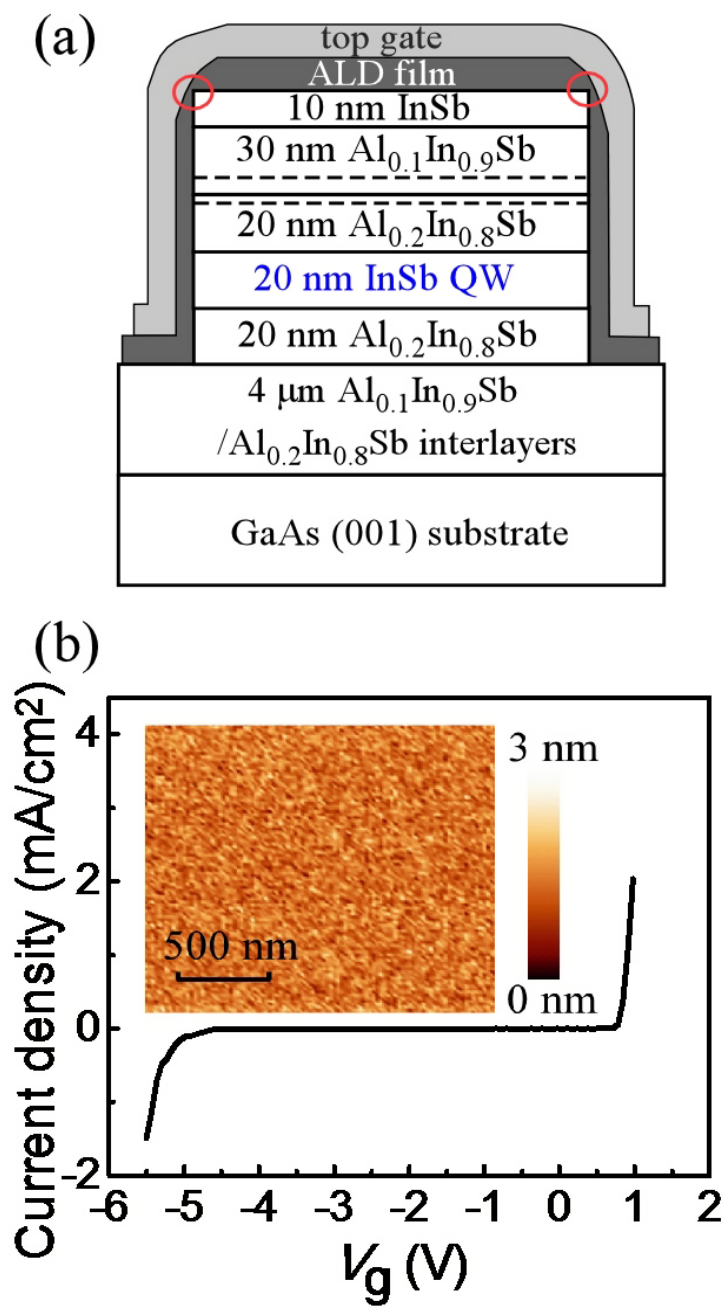

FIG. 1

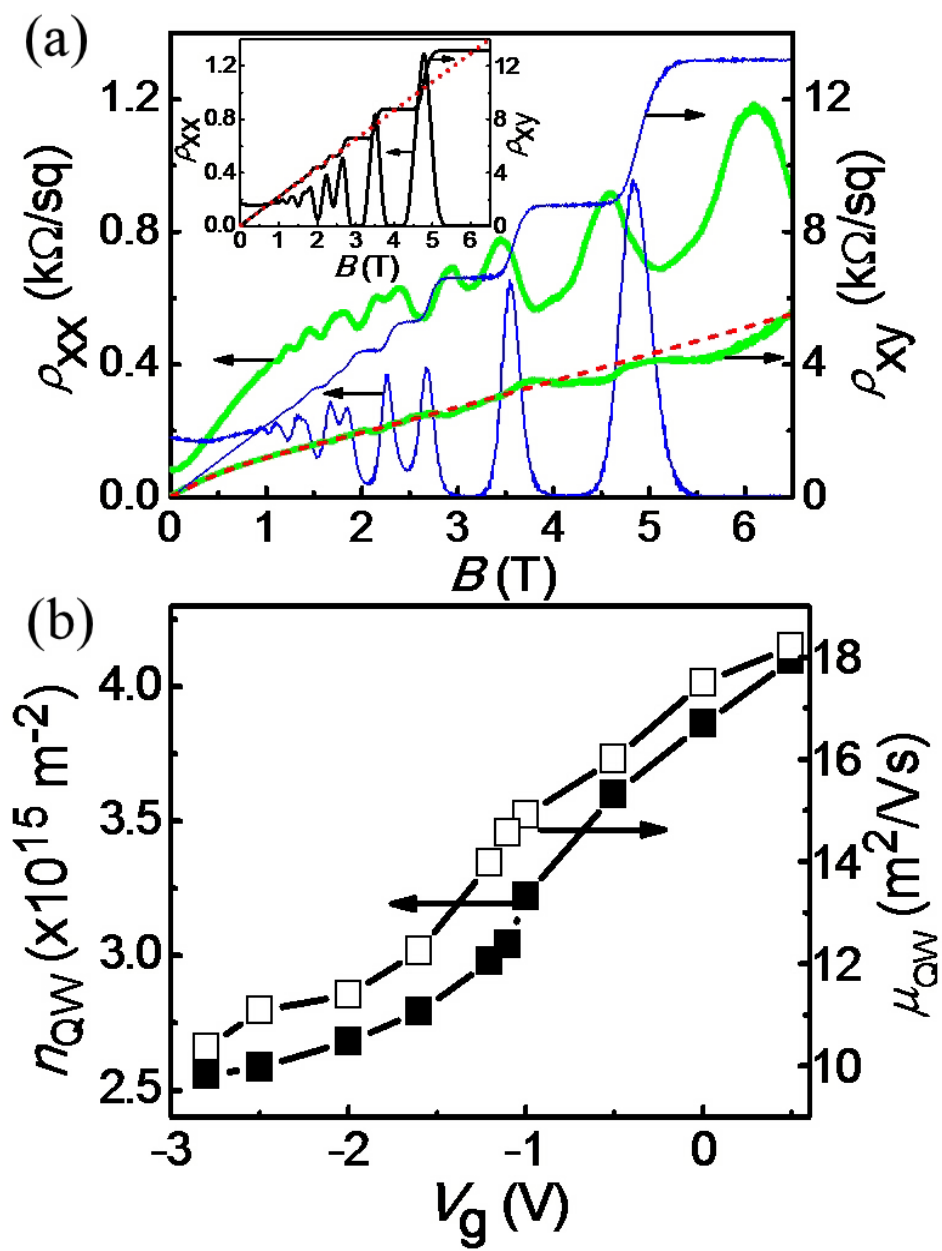

FIG. 2

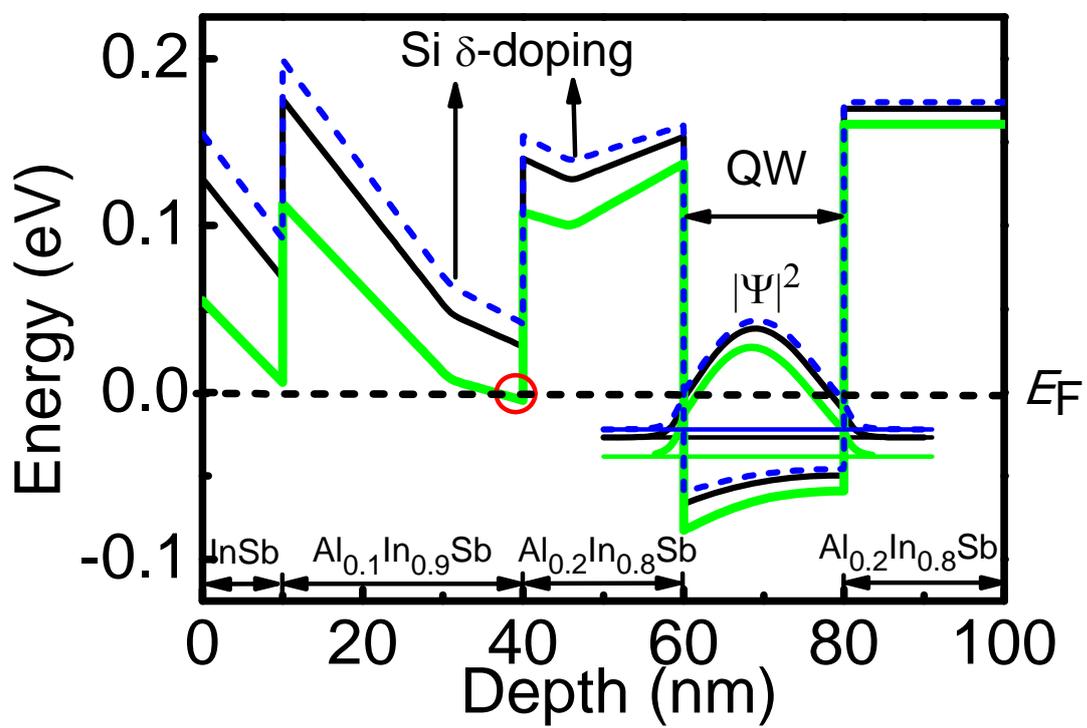

FIG. 3